\documentclass[5p]{elsarticle}

\usepackage[english]{babel}
\usepackage{indentfirst} 
\usepackage[latin1,utf8]{inputenc}
\usepackage{lmodern}
\usepackage[T1]{fontenc}
\usepackage{setspace}
\usepackage{hyperref}
\usepackage{color}

\usepackage{cellspace} 
\usepackage{amsmath,amsfonts,amssymb}

\usepackage{amsmath,amsfonts,amssymb}

\journal{Fusion Engineering and Design}

\begin{document}
\title{Neutron availability in the Complementary Experiments Hall \\ of the IFMIF-DONES facility}

 \author[1]{J.~Hirtz}
 \author[1]{A.~Letourneau}
 \author[1]{L.~Thulliez}
 \author[2]{A.~Ibarra}
 \author[3]{W.~Krolas}
 \author[3]{A.~Maj}
 
 \affiliation[1]{organization={IRFU, CEA, Université Paris-Saclay}, 
            city={Gif-sur-Yvette},
            postcode={91191},
            country={France}}
\affiliation[2]{organization={CIEMAT, Avda. Complutense 40}, 
            city={Madrid},
            postcode={28040},
            country={Spain}
}
\affiliation[3]{organization={Institute of Nuclear Physics PAN (IFJ PAN)}, 
            city={Krakow},
            country={Poland}
}

\begin{abstract}
 The IFMIF-DONES facility will be dedicated to the irradiation of structural materials planned for the use in future fusion reactors such as DEMO (Demonstration Fusion Power Plant).
 The potentialities of the IFMIF-DONES facility to complement its principal purpose by other experiments that would open the facility to other communities is addressed in this work.
 It concerns a study based on simulations to evaluate the neutronic performances of IFMIF-DONES in an hall dedicated to complementary experiments where neutrons can be transported.
 With the simple beam tube geometry of 4.5~cm entrance diameter studied in this work we have shown that a collimated fast-neutron beam of about 2~10$^{10}$~n/cm$^2$/s is available in the hall.
 Adding a moderator in the hall with neutron extraction lines would allow to get thermal neutron beams of about ~10$^{6}$-10$^{7}$~n/cm$^2$/s for dedicated experiments.
 The results show that IFMIF-DONES has the potentialities to be a medium-flux neutron facility for most of the neutron applications.
\end{abstract}

\maketitle

\section{Introduction}
\label{intro}

 Neutrons are unique tools to probe matter and its bulk properties.
 They can reveal what other techniques cannot see. Therefore, they provide a powerful tool for investigating the natural world and have long played a role at the forefront of modern scientific research.
 
 At the elementary and cosmological levels, neutrons provide very precise tests of the standard model of particle physics \cite{PPNS}. Moreover, the neutron lifetime is an important quantity that governs the primordial nucleosynthesis of elements a few seconds after the big bang. The neutron could also bring some of the answers on the origin of the matter-antimatter asymmetry in the universe and probe gravity at small scale \cite{ESS}. 
 
 In material and biological sciences, neutrons are unique to scrutinise both the structure and dynamics of atoms and molecules over a large range of distances and times. Compared to other standard techniques, as X-rays for example, neutrons scatter on nuclei instead of scattering on electrons offering a larger penetration in matter for most elements. Their interaction with hydrogen atoms allows very precise studies of organic materials as polymers, proteins, liquid crystals, etc.... Moreover, neutrons are very sensitive to magnetic fields allowing the study of magnetic materials.
 
 Neutrons have also a wide  range of applications in neutron radiography and tomography for various fields of science as life science, geology, archaeology but also for non-destructive testing of industrial components  \cite{tomo}. They play also an important role in nuclear medicine for radioisotope productions \cite{medic}.
 
 The IFMIF-DONES (International Fusion Materials Irradiation Facility - DEMO Oriented NEutron Source) \cite{IFMIF-DONES} is a project of an accelerator based deuterium-lithium neutron source for irradiation of fusion reactor materials. A 125 mA beam of 40 MeV deuterons will strike a liquid Li target and produce an unprecedented high flux of neutrons with energies up to 50~MeV. It is part of an international effort to pave the road toward fusion energy. 
 Fusion material irradiations will take place in the high-flux test module (HFTM) located in the test cell, the central confinement to envelop the end section of the deuteron accelerator and the target. The HFTM will be placed close to the Lithium target to benefit from the maximum neutron flux amounting to about $10^{14}$~n/cm$^2$/s~\cite{ifmif}.
 A large part of the neutrons will interact with the materials to be irradiated but most of them will just scatter-off or pass through the irradiation module and will be lost in the walls of the test cell bio-shield.

 An effort, in the framework of the IFMIF-DONES Preparatory Phase \cite{DONESPrep}, has been conducted to study the possibility of using neutrons for other experiments in a dedicated hall \cite{Krolas}.
 This hall will be adjacent to the test cell bio-shield, a neutron beam line passing through the bio-shield and allowing neutrons to be transported from the test cell to the hall.
 
 \begin{figure*}[t]
\begin{minipage}{2\columnwidth}
\centering
\includegraphics[width=0.8\columnwidth]{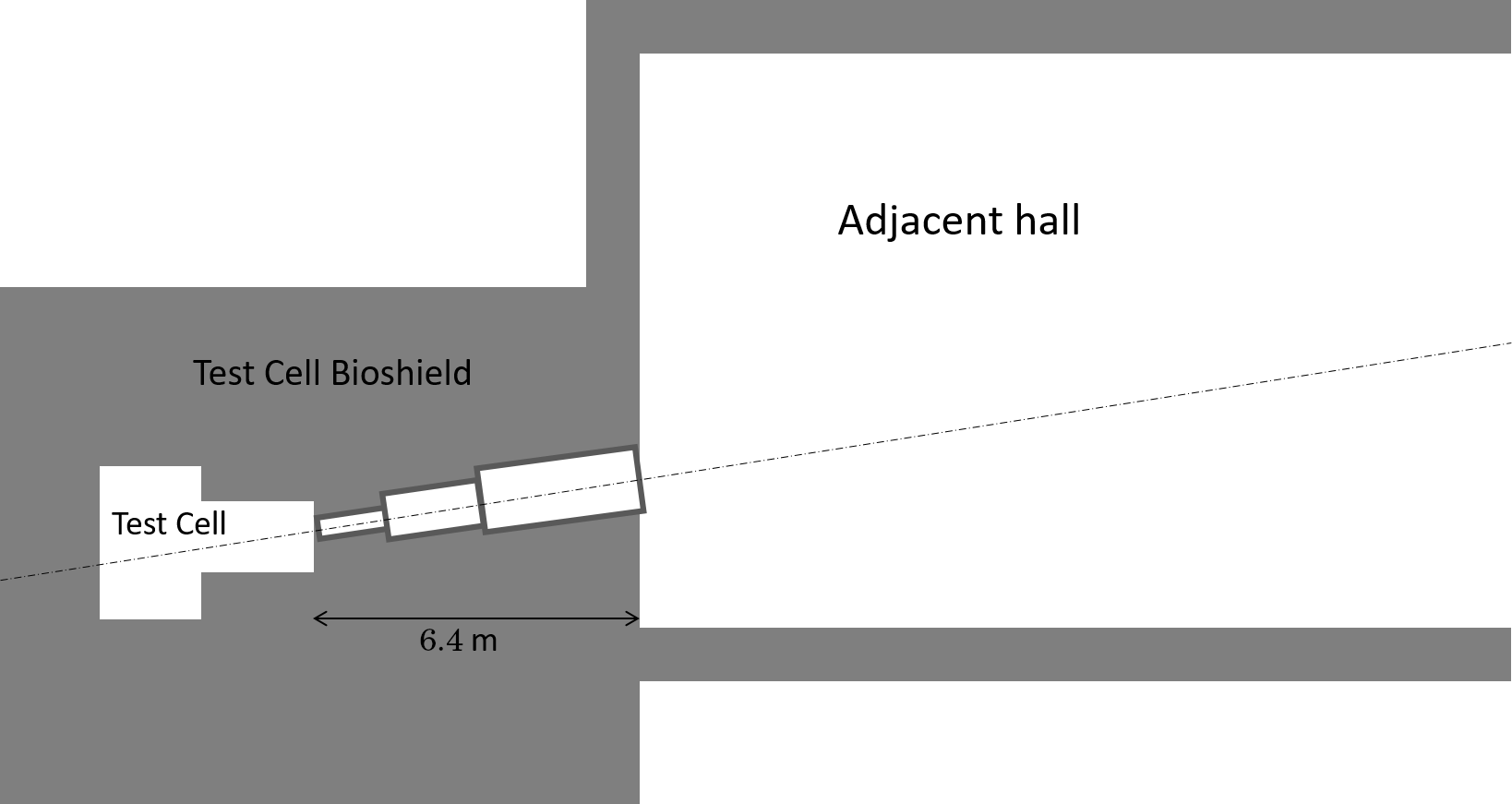}
\caption{\label{plan} Schematic view of the test cell, the complementary experiments hall and the beam line defined in this work (not to scale). The axis of the deuteron and neutron beam line is indicated by the dotted-dashed line. The entrance and exit diameters of the beam line are 4.5~cm and 9~cm, respectively (see text for details). }
\end{minipage}
\end{figure*}
 
The paper reports the simulation work performed to characterise the neutron beams that will be available in the complementary experiments hall (see \autoref{plan}). It is organised as follows.
In \autoref{soft}, we present the methodology and software used in our study.
Next, the neutron beam characteristics that would be available in the hall are given in \autoref{chara} for unmoderated neutrons together with background.
And finally, the various fluxes which could be available in the hall as a function of the setup chosen are displayed in \autoref{moderation}.

\section{Methodology and software tools}
\label{soft}

The global methodology followed in this work relies on Monte-Carlo simulations to evaluate and optimise the neutronic performances of the installation.

Simulations are based on Geant4.10.7~\cite{geant4} using the TOUCANS (TOolkit for a Unified Compact Accelerator Neutron Sources design) toolkit~\cite{toucans} and MCNP6.2 (Monte Carlo N-Particle Transport Code System Version 6.2)~\cite{mcnp}.
 TOUCANS is a modular code using Geant4 C++ libraries.
 It is developed at CEA Saclay since 2019 to design Compact Accelerator Neutron Sources (CANS) and to evaluate their neutronic performances.
 Geant4 (GEometry ANd Tracking) is a framework developed at CERN for the simulation of the passage of particles through matter by the Monte-Carlo method.
 Monte Carlo N-Particle or MCNP is a general-purpose Monte Carlo radiation transport code designed to track many particle types over broad ranges of energies.
 It is registered trademark owned by Triad National security, manager and operator of Los-Alamos National Laboratory (LANL).

 The neutron transport in Geant4.10.7 at low energy was validated on experimental measurements (see ref.~\cite{sonate} and references therein).
 More recently, it has been benchmarked for cold moderators with Tripoli-4$^{\mbox{\scriptsize{\textregistered}}}$ \cite{tripoli}, a transport Monte Carlo code developed at CEA for reactor physics with depletion, criticality and safety.
 The benchmark showed a perfect agreement between the two codes \cite{toucans}.

 The simulations carried out with Geant4 were based on the physics list QGSP\_BIC\_AllHP.

 The double differential neutron spectrum inside the test cell used as input for Geant4 simulations was calculated by means of the MCNP-6 code using the DONES model ``mdl9.2.1'', which includes the updated HFTM and test cell configuration.
 In that simulation, the package McDeLicious-17~\cite{McDeLicious} was capitalised on to simulate the D-Li reactions.
 The neutron FENDL-3.1d library~\cite{fendl} was employed.
 The results were normalised to a 125~mA beam of 40~MeV deuterons impinging the lithium target.

\begin{figure}[h]
\hspace{3mm}\includegraphics[width=.90\columnwidth]{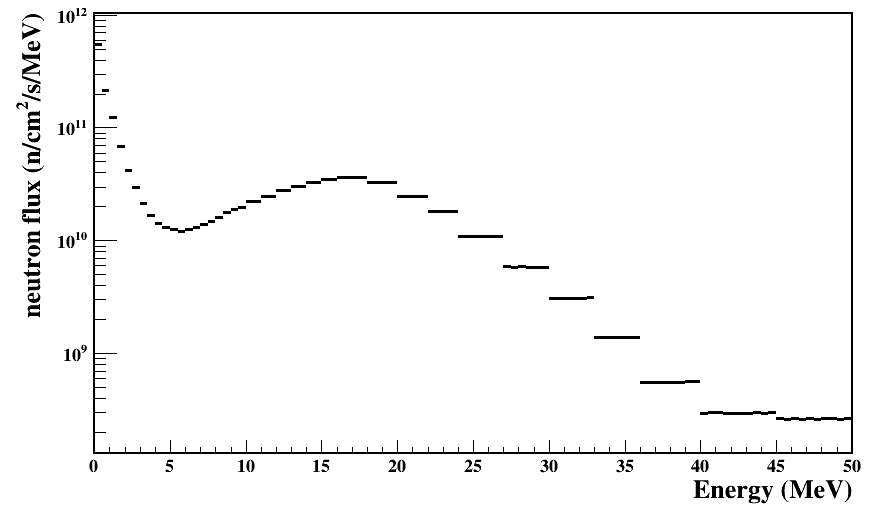}
\caption{\label{incoming} Neutron flux at the entrance of the inlet in the test cell. The plateaux observed at high energy is due to the binning in our input file.}
\end{figure}

 The source term for the Geant4 simulations was extracted from these simulations by fixing a circle with a radius of 2.25~cm at the entrance of the inlet.
 In this configuration, the neutron current at the entrance of the beam tube is 1.835~10$^{13}$~n/s (1.15~10$^{12}$~n/s/cm$^2$) as shown in \autoref{incoming}.
 This source term is available in the complementary material of this article.

 The number of simulated histories in MCNP-6 simulations was 2~10$^{10}$ to get sufficient statistics.
 In the case of Geant4, the number of simulated histories was highly variable as a function of observables studied.
 This number varies between $10^6$ for the most accessible observables and $10^9$ when statistic was needed.

\section{Unmoderated neutron beam in the hall}
\label{chara}
\subsection{Neutron characteristics}

 The case of reference chosen in this work is a beam line of conical shape to transport neutrons from the test cell to the complementary experiments hall (see \autoref{plan}).
 The beam tube is constituted of three cylindrical iron tubes of 4.5~cm, 6~cm and 9~cm in diameters with respective length of 1~m, 1.7~m and 3.7~m.
 This shape allows to follow the natural dispersion of the neutrons to maximize the neutron flux in the hall but a cylindrical shape could be adopted if a better collimation is needed.
 It is aligned on the deuteron beam axis and tilted by 9° with respect to the test cell bio-shield wall. In the configuration studied here, the bio-shield thickness is 6.4~m.
 
 \begin{figure}
\centering
\includegraphics[width=\columnwidth]{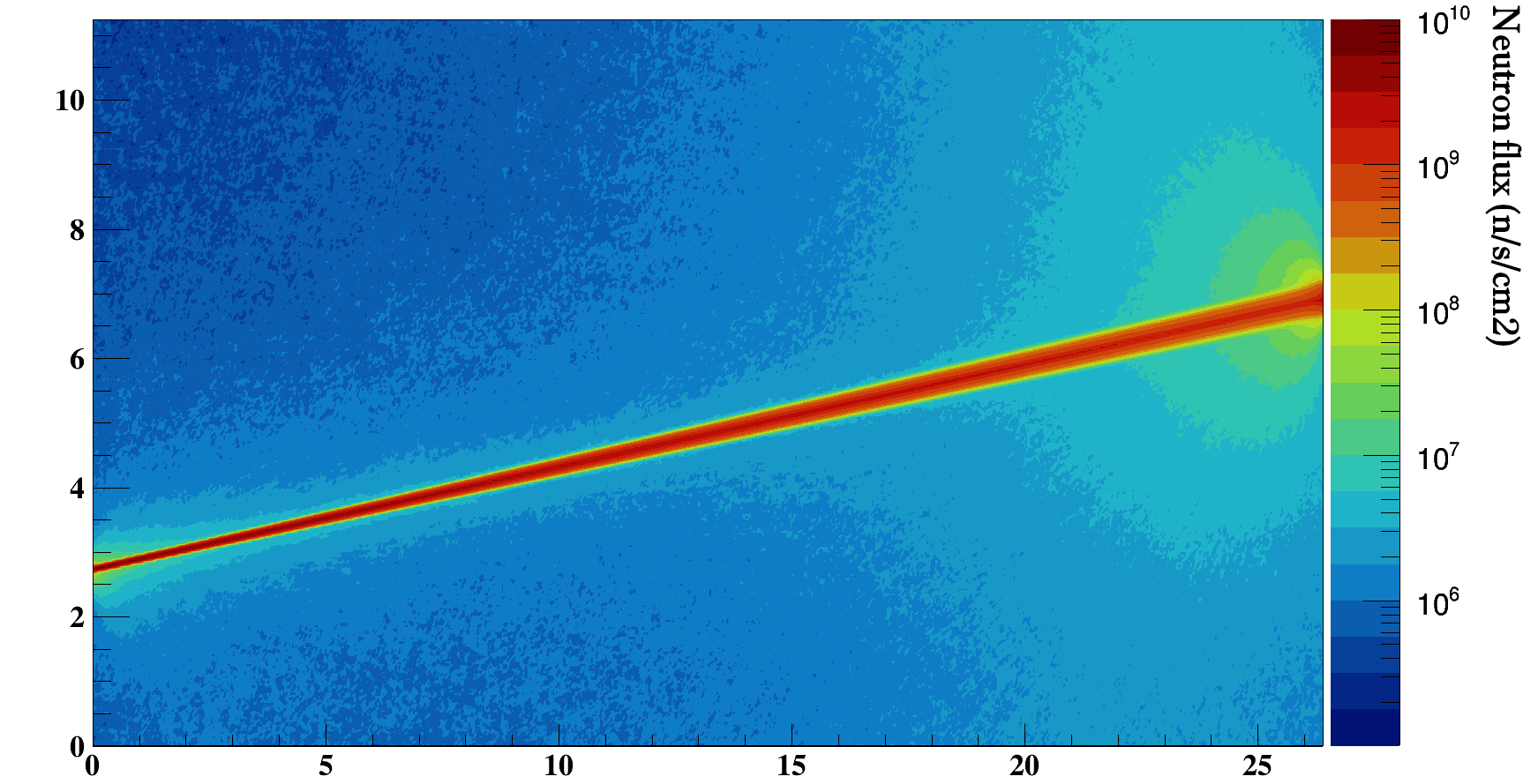}
\caption{\label{neutMid} Neutron flux map in the complementary experiments hall for unmoderated neutrons at beam level.}
\end{figure}
 
\begin{figure}
\centering
\includegraphics[width=0.9\columnwidth]{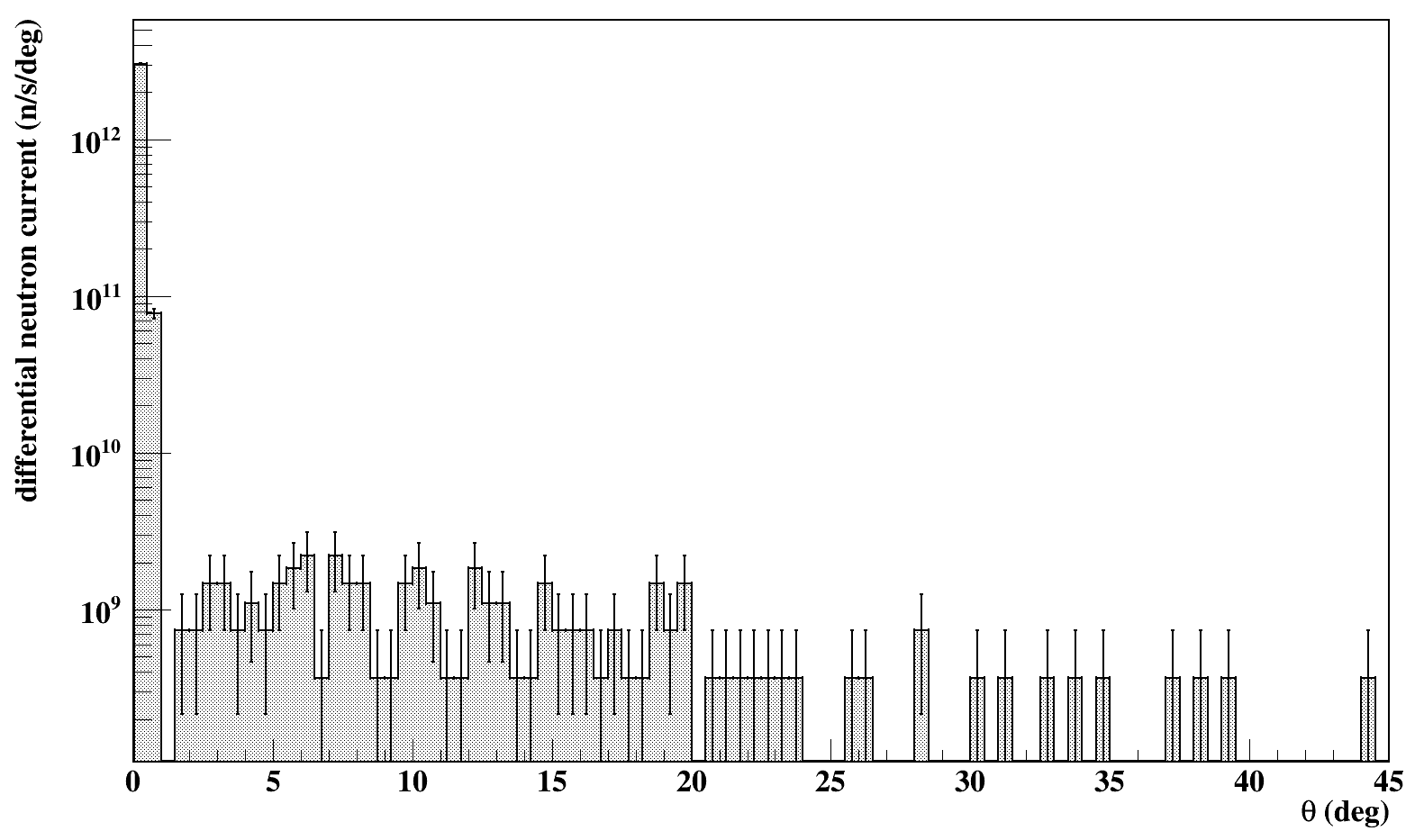}
\caption{\label{angle} Divergence of the neutrons at the entrance of the hall with respect to the neutron beam axis.}
\end{figure}

\begin{figure}
\centering
\includegraphics[width=\columnwidth]{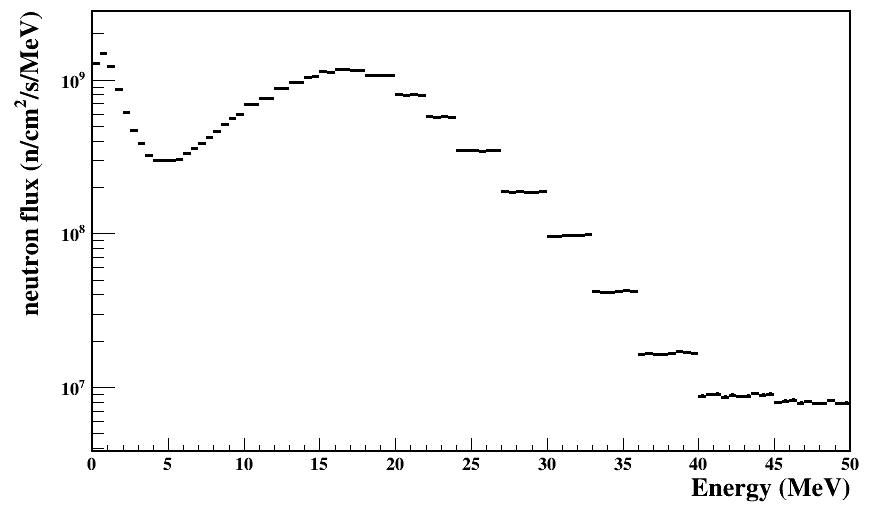}
\caption{\label{InOut} Unmoderated neutron flux at the entrance of the hall.}
\end{figure}

\begin{figure}
\centering
\hspace*{3mm}\includegraphics[width=.9\columnwidth]{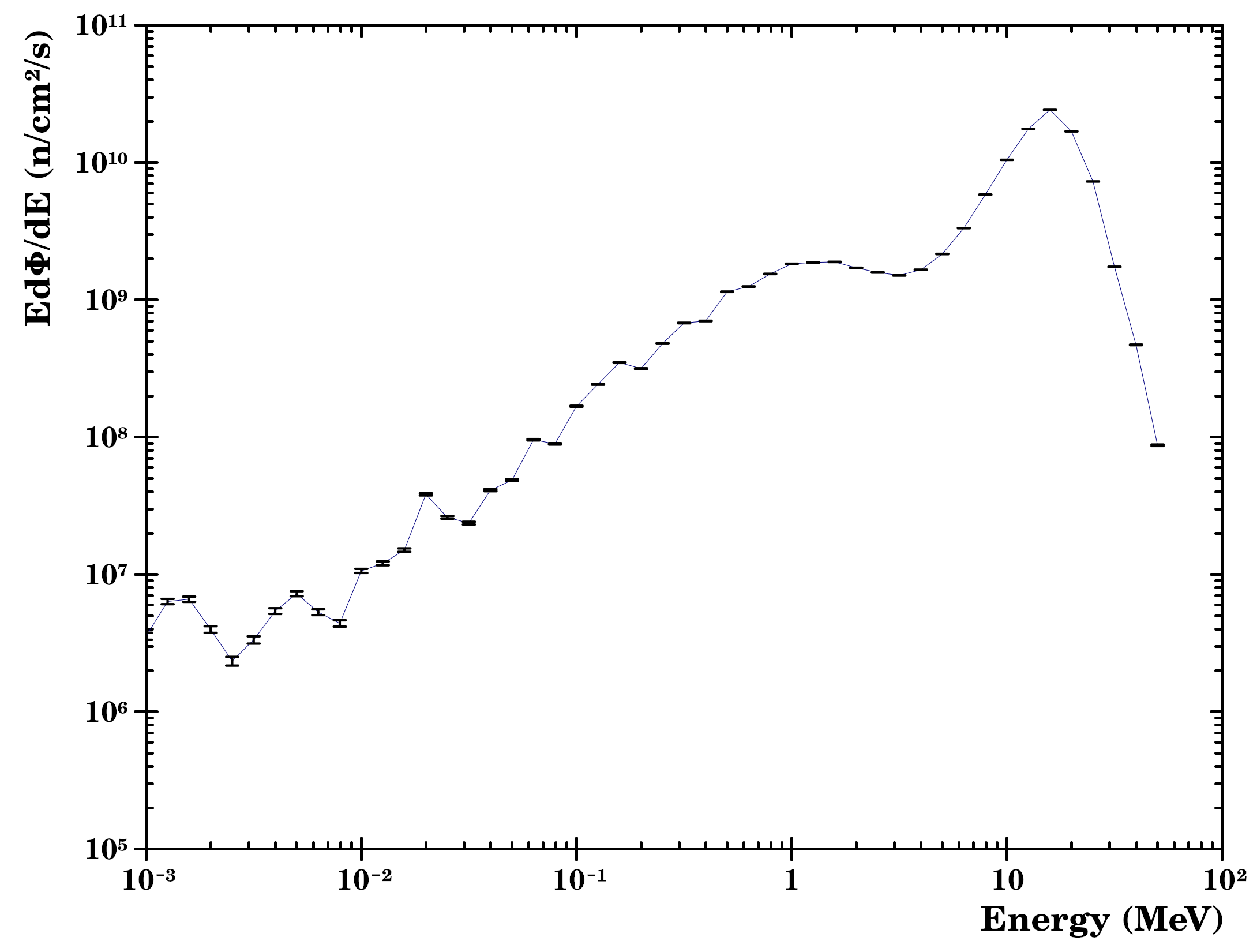}
\caption{\label{lethargy} Same as \autoref{InOut} in lethargy unit.}
\end{figure}

 The neutron beam characteristics at the entrance of the hall are shown in \autoref{neutMid}, \ref{angle}, \ref{InOut}, and \ref{lethargy}.
 The beam is very well collimated with a global divergence of less than 1~degree (see \autoref{angle}).
 However, there is no beam structure and therefore, Time of Flight (ToF) technique cannot be used.
 This is compensated by a very high neutron flux in comparison to other facilities.
 A tabulation of the neutron flux at the entrance of the hall as a function of the energy and the angle of emission is available in the complementary material.

 With this configuration of beam tube, the neutron flux emerging in the hall amounts to a total of 2~10$^{10}$~n/cm$^2$/s with a majority (99.5\%) of fast neutrons (E$_n$ > 100~keV) to be compared with the neutron flux in the test cell at the entrance of the collimator ($\sim$10$^{12}$~n/cm$^2$/s).
 The energy distribution remains the same along the collimator for the energies above 5~MeV with a scaling effect of about 8\% due to the opening of the solid angle and the absorption of the neutrons in the first part of the collimator (see \autoref{InOut}).
 However, the spectrum is modified below 5~MeV with a selection in favour of the highest energies while low energy neutrons are strongly reduced.
 3\% of neutrons with E$_n$ < 5~MeV rich the hall while 15\% for E$_n$ > 5~MeV.
 The lost neutrons result in an energy deposition in the collimator and the nearby concrete.
 It affects mainly the first module of the collimator with a maximal energy deposition of about 5~mW/cm$^3$, including the activation of material, to be compared with the 50~mW/cm$^3$ in this region due to the near vicinity of the test cell.

\begin{figure}
\centering
\includegraphics[width=\columnwidth]{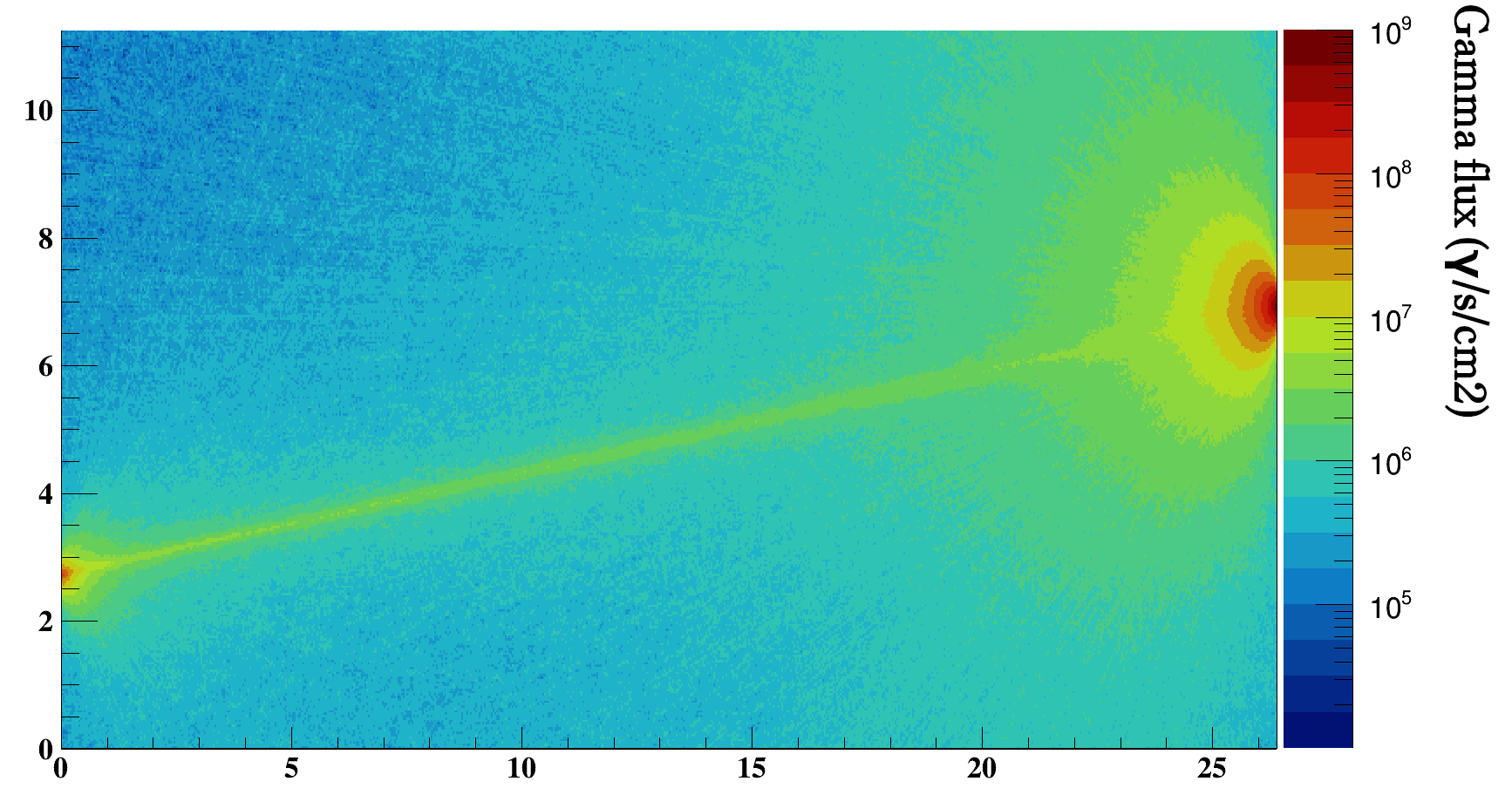}
\caption{\label{gammaMid} Gamma-ray flux map in the complementary experiments hall due to the neutron beam at the beam level.}
\end{figure}

\subsection{Beam dump}

 In the meanwhile, the very well collimated and energetic neutron beam creates a ``hot spot'' on the opposite wall of the hall and generates secondary gammas and neutrons (see \autoref{neutMid} and \autoref{gammaMid}).
 Both the fast neutron and the gamma-ray fluxes in the hall call for the moderation of the neutron beam inside the hall in order to avoid radioprotection issues in the nearby rooms and corridors.
 It is also necessary to reduce the neutron and gamma-ray background for experiments.

\begin{figure}[h]
\centering
\includegraphics[width=0.45\columnwidth]{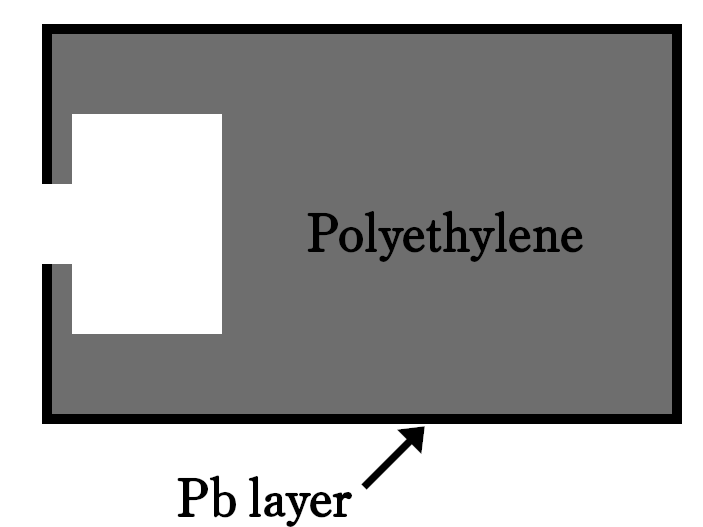}\includegraphics[width=0.45\columnwidth]{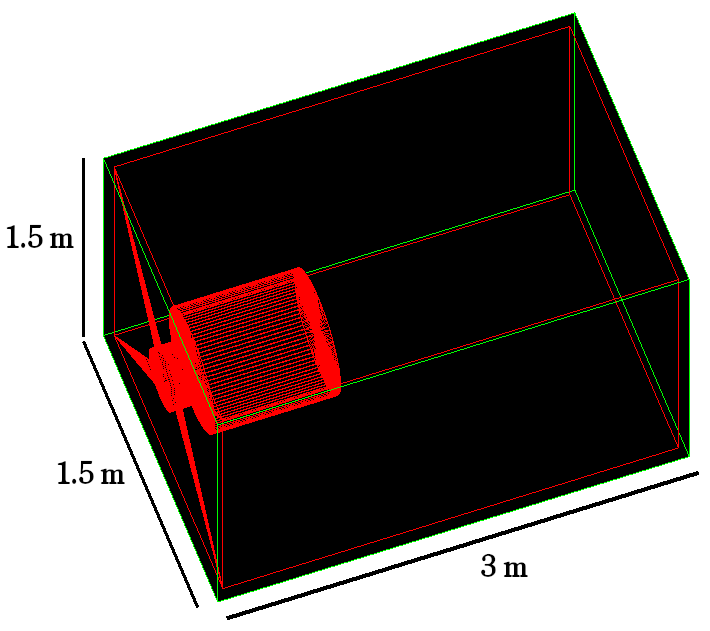}
\caption{\label{schema} Schematic (left) and 3D view (right) of the beam dump.}
\end{figure}

 In order to avoid the creation of a hot spot regardless to the presence or not of material along the neutron beam line, a beam dump design has been studied.
 Such beam dumps exist in different laboratories hosting neutron beams.
 The main characteristics of those beam dumps have been reviewed and conventional solutions adopted at other facilities have been considered for this study.

\begin{figure}[h]
\centering
\includegraphics[width=\columnwidth]{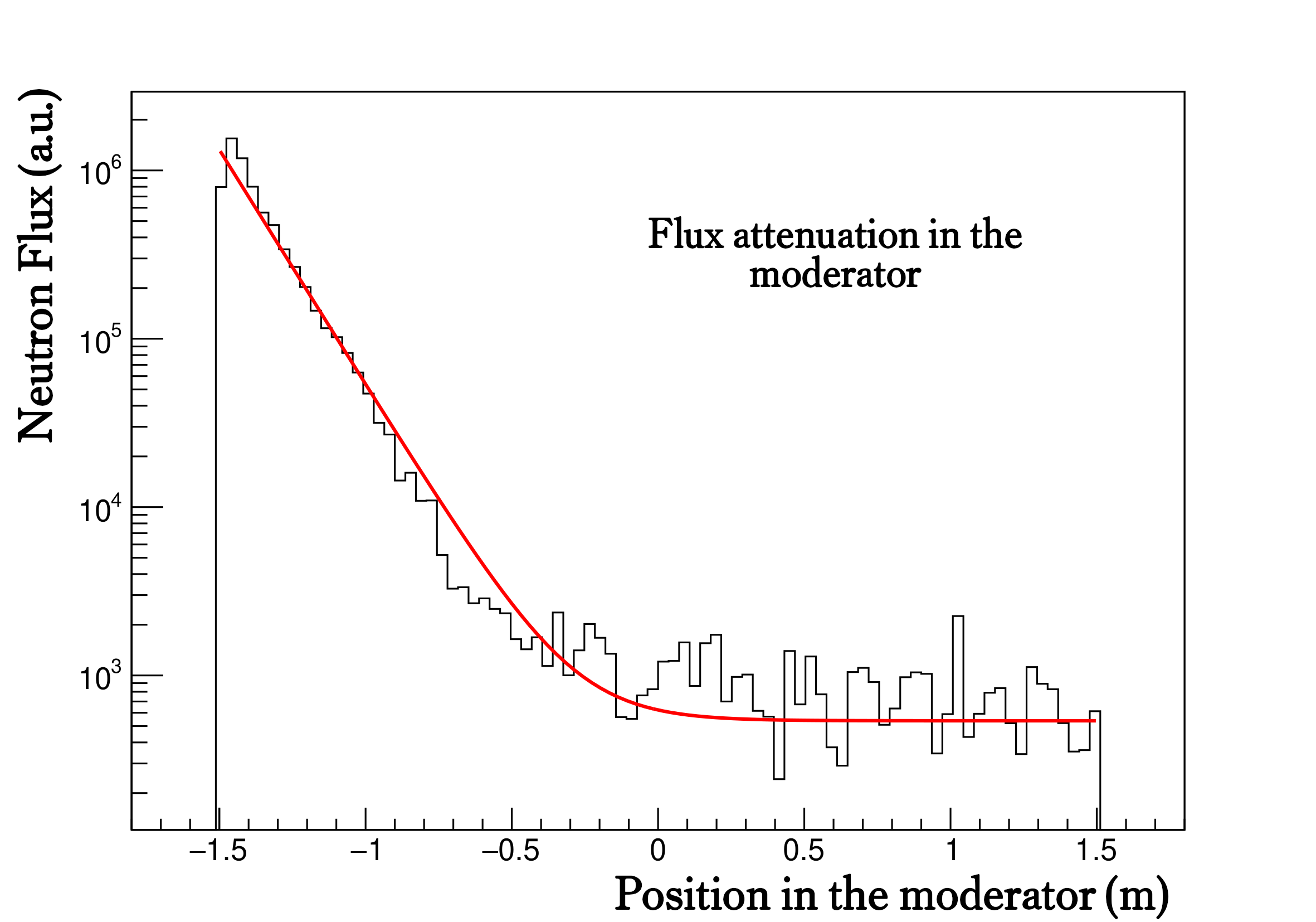}
\caption{\label{atte2} Neutron flux attenuation in the beam dump.}
\end{figure}

\begin{table*}
\begin{minipage}{2\columnwidth}
\begin{center}
\caption{\label{tableGeo} Different trap geometries for the beam dump (see also \autoref{BeamDump}).}
\begin{tabular}{|m{0.1\textwidth}<{\centering}|m{0.53\textwidth}|b{0.2\textwidth}<{\centering}|}
\hline
Trap ID & \begin{center} Description of the trap geometry \end{center} & Relative backward neutron flux (\%) \\
\hline
a & A 1~m long cylindrical hole with a radius of 40~cm & $ 1.69\pm 0.18$ \\
\hline
b & A 20~cm long cylindrical hole with a radius of 20~cm followed by a 80~cm long cylindrical hole with a radius of 40~cm & $1.25 \pm 0.12$ \\
\hline
c & Same as b) with a 50~cm long cone of polyethylene with a base radius of 30~cm & $1.56\pm0.16$ \\
\hline
d & Same as c) with a cylindrical hole with a radius of 5~cm & $1.17\pm0.11$ \\
\hline
e & Same as b) with a 50~cm long cylinder of polyethylene with a radius of 10~cm & $1.93 \pm 0.22$ \\
\hline
f & Same as e) with an inclination of the polyethylene cylinder surface by 45$^\circ$ & $2.10 \pm 0.25$ \\
\hline
g & Same as e) with a cylindrical hole with a radius of 5~cm & $1.32\pm 0.12$ \\
\hline
h & Same as d) with an additional conical hole with a base radius and a height of 20~cm & $1.69 \pm 0.18$ \\
\hline
\end{tabular}
\end{center}
\end{minipage}
\end{table*}
 
 The position retained for the beam dump is 20~m from the exit of the beam tube, close to the opposite wall.
 This allows for a large space in front of the beam dump for experiments.

 The beam dump must meet three objectives.
 First, it must stop the fast neutron beam in order to avoid a hot spot on the backward wall.
 Second, it must limit the backward reflection of thermal and fast neutrons.
 Finally, it must limit the gamma-ray flux in the hall due to the neutron moderation.

 To fit within these objectives, we opted for a cubic (1.5~m $\times$ 1.5~m $\times$ 3 m) polyethylene moderator along the neutron beam line with a 5~cm thick layer of lead at its surface (see \autoref{schema}).
 The lead layer at the surface of the beam dump strongly reduces the gamma emission from the beam dump.
 
\begin{figure}[h]
\centering
\includegraphics[width=0.9\columnwidth]{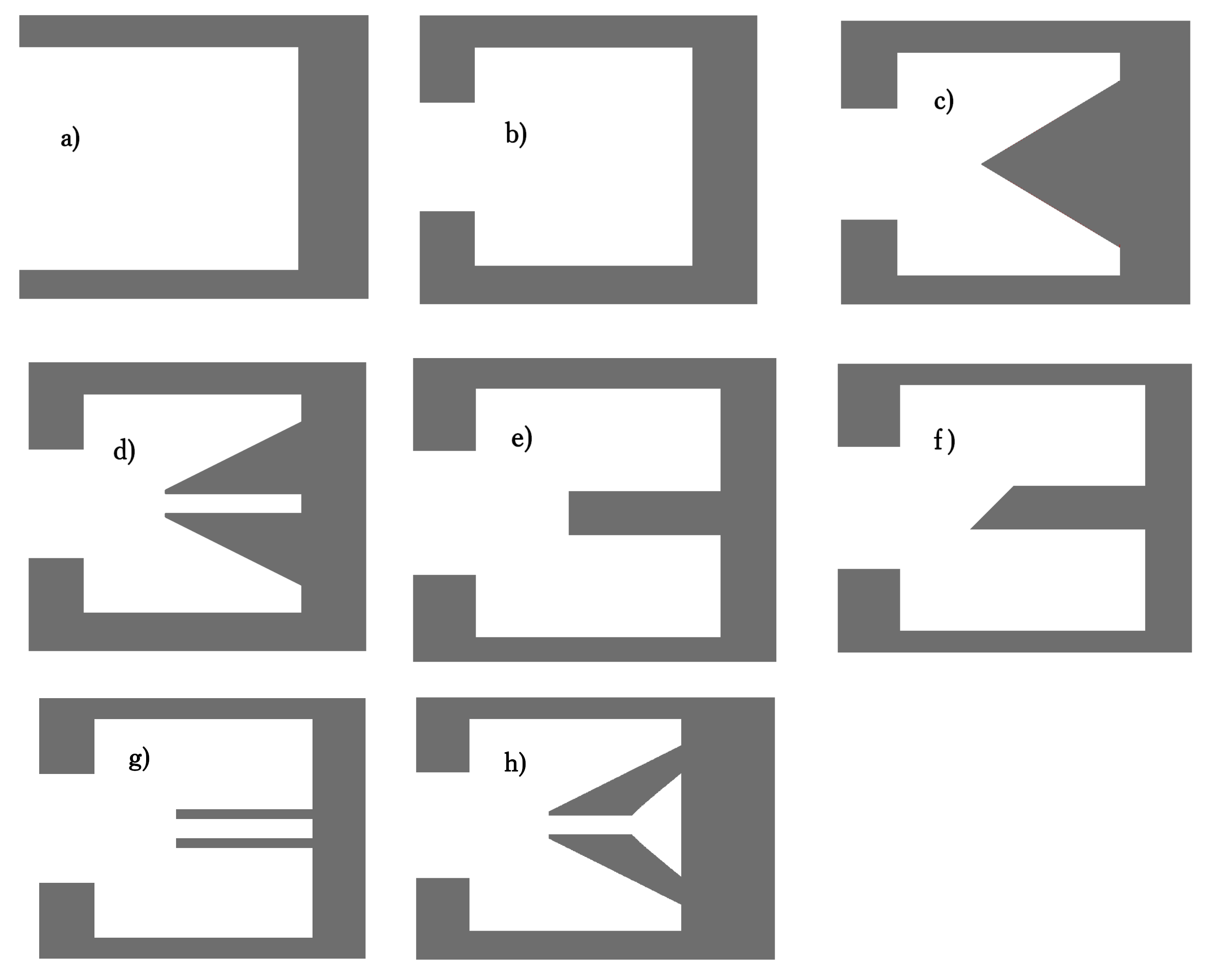}
\caption{\label{BeamDump} Schematic slices of the trap geometries for the beam dump.}
\end{figure}

 The thickness of the polyethylene was chosen based on the attenuation in the moderator (\autoref{atte2}).
 Regarding the neutron beam spectrum, after 1.5~m of moderation, the neutron flux in the moderator is at the same order of magnitude than in the room due to scattered neutrons (see \autoref{neutMid}).
 We then consider an extra 50~cm of security margin for a total thickness of 2~m (195~cm of polyethylene + 5~cm of lead).

\begin{figure*}
\begin{minipage}{2\columnwidth}
\centering
\hspace{3mm}\includegraphics[height=0.25\columnwidth ,width=0.475\columnwidth]{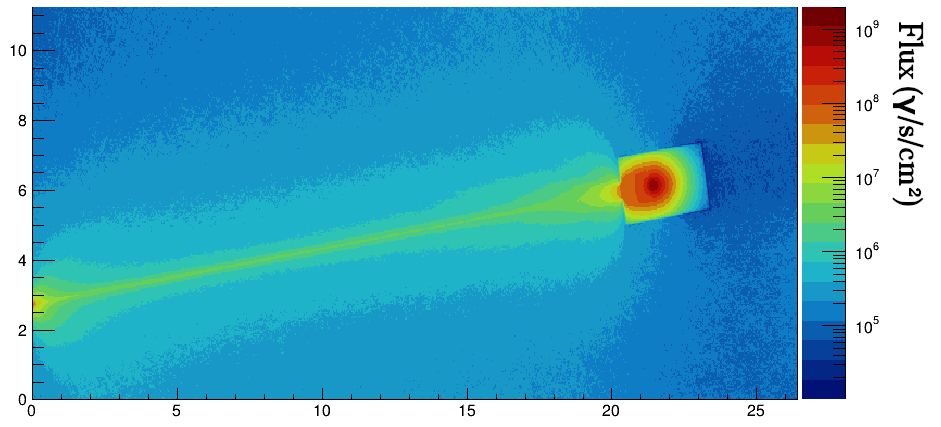}\includegraphics[height=0.25\columnwidth ,width=0.48\columnwidth]{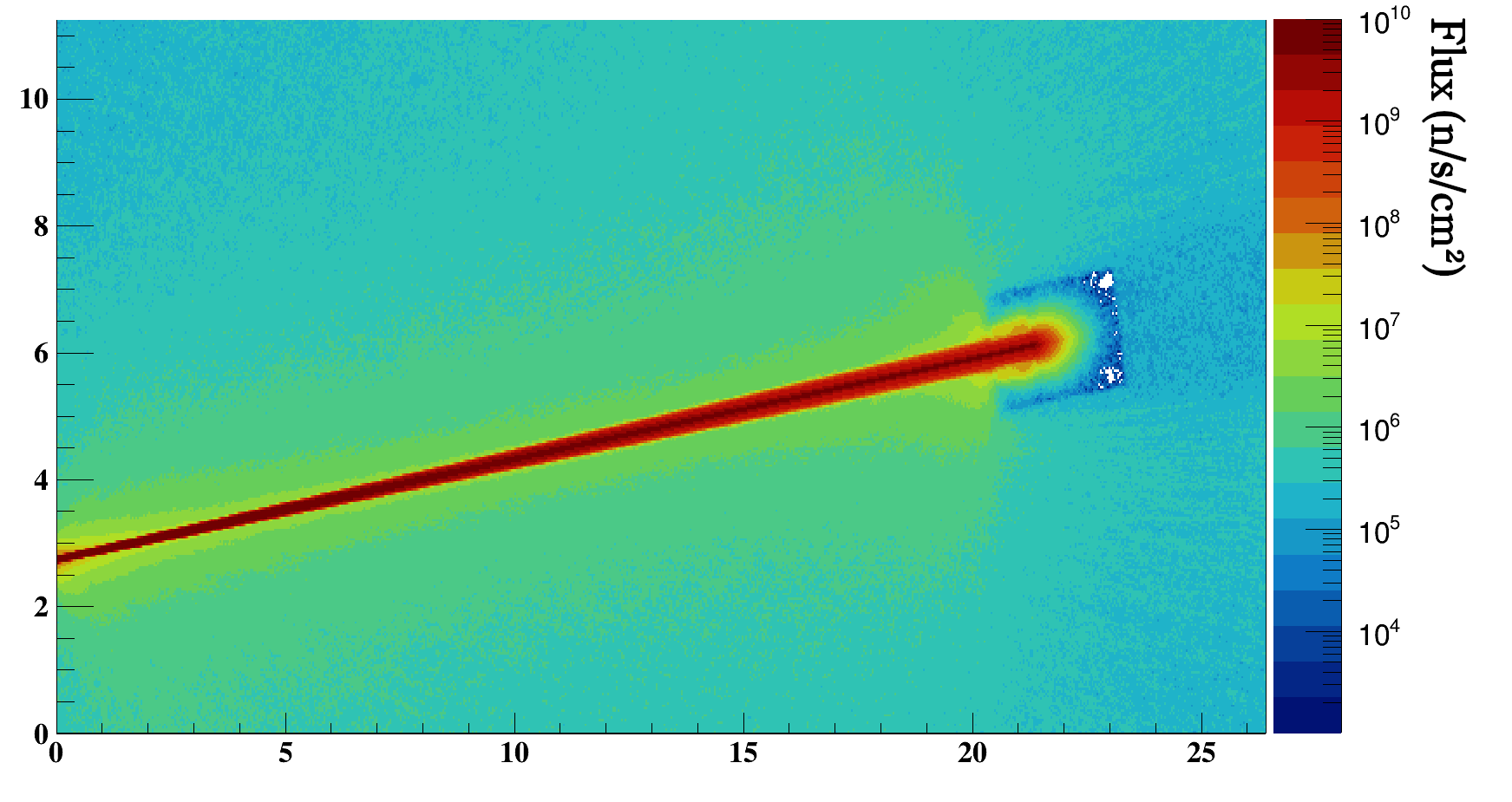}\\
\includegraphics[height=0.25\columnwidth ,width=0.48\columnwidth]{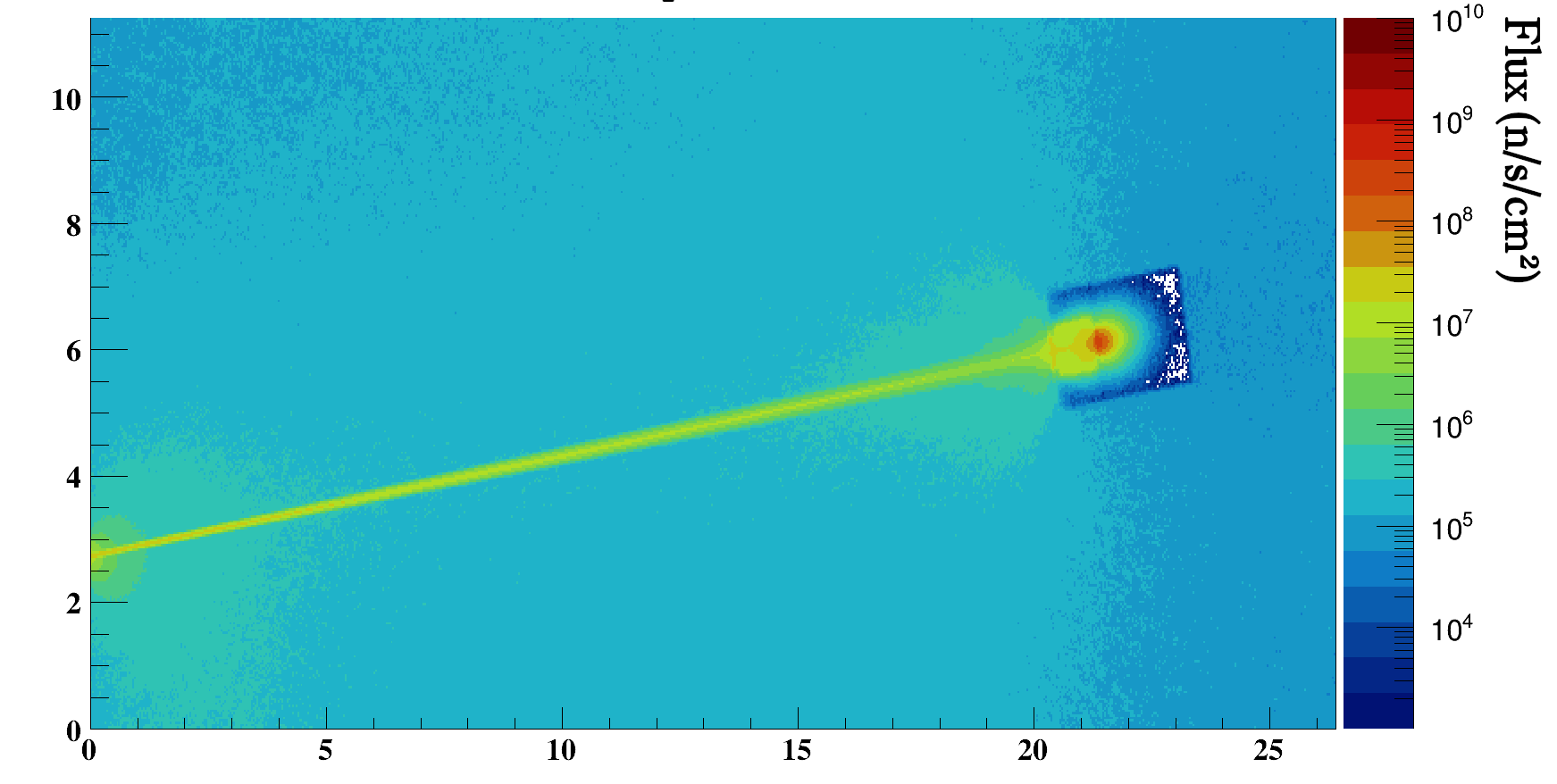}\includegraphics[height=0.25\columnwidth ,width=0.48\columnwidth]{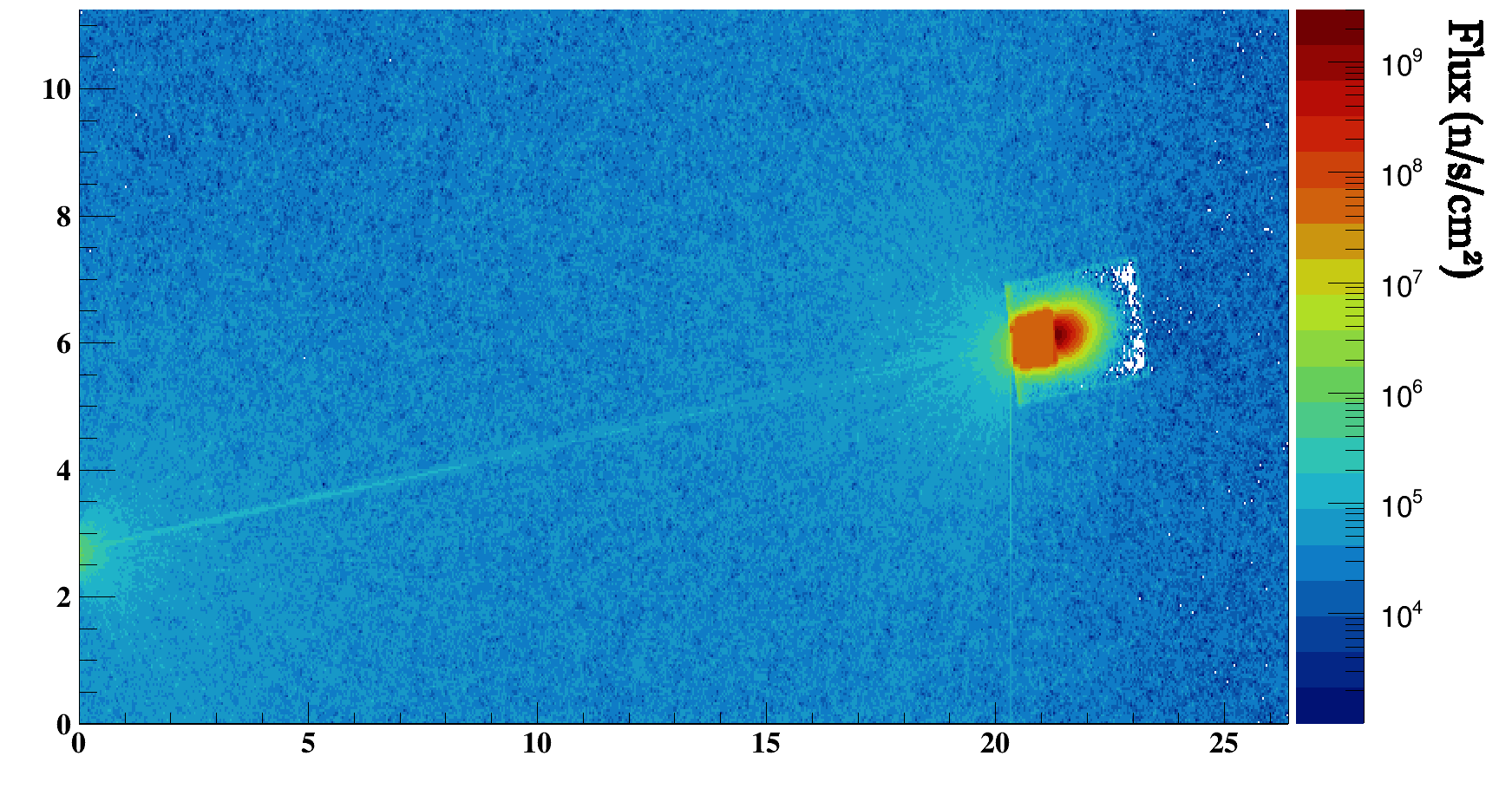}
\caption{\label{WithBeamDump} Neutron and gamma-ray fluxes in the hall in presence of a beam dump. Top left: gamma-rays. Top right: fast neutrons. Bottom left: epithermal neutrons. Bottom right: thermal neutrons.}
\end{minipage}
\end{figure*}
 
 The backward neutron and gamma fluxes are reduced thanks to a ``trap'' at the entrance of the beam dump.
 Various shapes have been tested for this trap as shown in \autoref{BeamDump} and \autoref{tableGeo}, and different lengths have been tested for the different elements of the trap.
 However, none of the complex structures tested improved significantly the backward fluxes.
 Here, the trap efficiency can be summarised as: ``the deepest, the best''.
 We then have chosen the (b) option in \autoref{tableGeo}, which consist in an 80~cm long cylinder of radius 40~cm with another 20~cm long cylinder acting as a door with a radius of 20~cm corresponding to the dispersion of the neutron beam at the location of the beam dump, i.e., 20~m from the exit of the beam tube.
 One can mention that the (d) structure shows a slightly better efficiency than the structure retained (b).
 However, the uncertainties do not allow to conclude with certitude and the reduction of the backward neutron flux is too small to be considered as a real improvement of the setup.
 With a total depth of 1~m (= 20~cm + 80~cm), the backward neutron flux represents 2.2\% of the entering flux (see flux maps for details).
 
 In \autoref{WithBeamDump} are displayed the gamma and thermal (E$_n<400$~meV), epithermal ($400~$meV$<$E$_n<100~$keV), and fast (E$_n>100~$keV) neutron fluxes in presence of the beam dump.
 It clearly demonstrates the efficiency of the design in stopping the fast neutron beam as well as limiting the backward emission of thermal neutrons and gamma-rays.

\section{Moderated neutrons}
\label{moderation}

 As seen in the previous section IFMIF-DONES will deliver energetic neutrons beam, but most of neutron applications require moderated neutrons, either thermal or cold neutrons.
 In this section we study the possibility of having moderated neutrons by placing a moderator at the exit of the neutron beam line, in the hall. This solution has the advantage to be very flexible and offers the possibility to design specific moderators for different uses with specific extraction lines and without drawback on the energy spectrum in the test cell.
 
\begin{figure}
\centering
\includegraphics[width=\columnwidth]{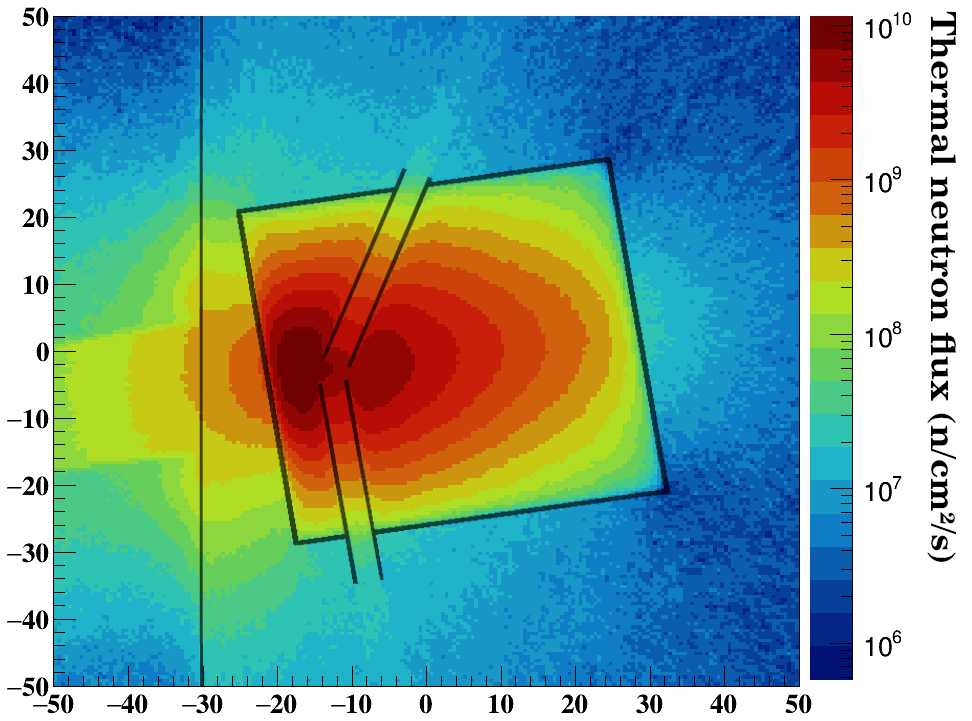}
\caption{\label{extra} Simulated thermal neutron flux inside the reference moderator setup (50$\times$50$\times$50~cm$^3$ polyethylene block with two neutron extraction lines tilted at 90° and 60° with respect to the beam axis) located at 10~cm from the exit of the beam tube (the concrete wall of the test cell bio-shield is indicated by the vertical line).}
\end{figure}

 We have studied various thermal moderators.
 In the following, we focus on what we call the reference setup: a 50$\times$50$\times$50~cm$^3$ polyethylene block located 10~cm from the exit of the beam tube as shown in \autoref{extra}, with 2 extraction lines.

\begin{figure}
\centering
\includegraphics[width=\columnwidth]{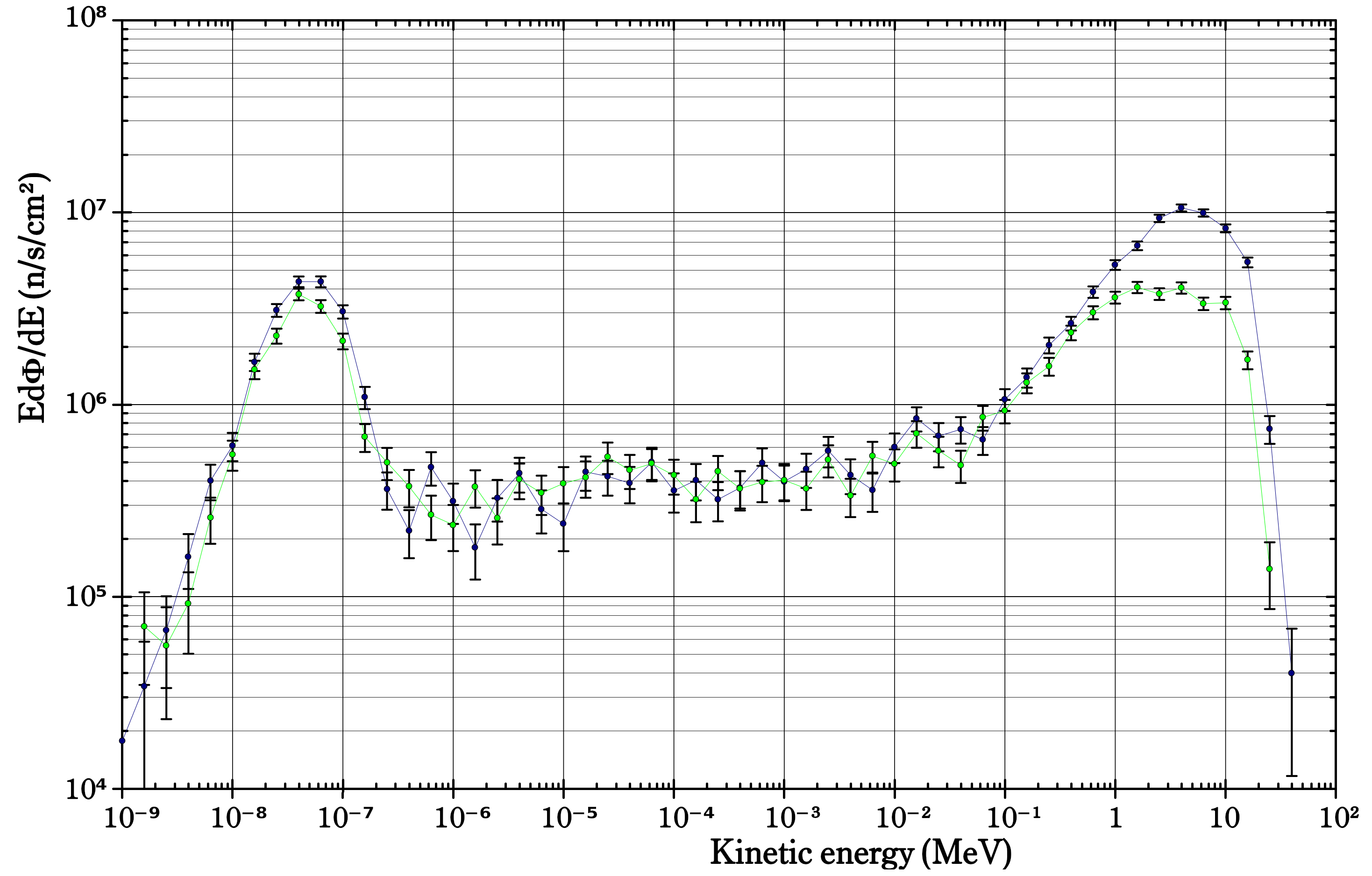}
\caption{\label{2021} Neutron flux distribution (in unit of lethargy) obtained at the exit of the extraction lines for the reference moderator setup (see Fig.\autoref{extra}). Blue: 90$^\circ$ extraction line. Green: 60$^\circ$ extraction line.}
\end{figure}

 The extraction lines are 30~cm long (see in \autoref{extra}) and with inner diameter of 3~cm.
 One is tilted by 90° with respect to the beam axis and the other one is tilted by 60°.
 They are positioned 10~cm away from the entrance of the neutron beam where the density of thermal neutrons is maximal.
 As shown in \autoref{2021}, the thermal neutron flux extracted is 8.7~10$^6$~n/cm$^2$/s for the 90$^\circ$ tilted line with respect to the neutron beam axis and 1.1~10$^7$~n/cm$^2$/s for the 60$^\circ$ tilted line.
 Additionally, the thermal to total ratio amounts to 25.1\% and 19.7\%, for a total neutron flux of 3.5~10$^7$~n/cm$^2$/s and 5.7~10$^7$~n/cm$^2$/s, respectively.

\begin{figure}
\centering
\includegraphics[width=.9\columnwidth]{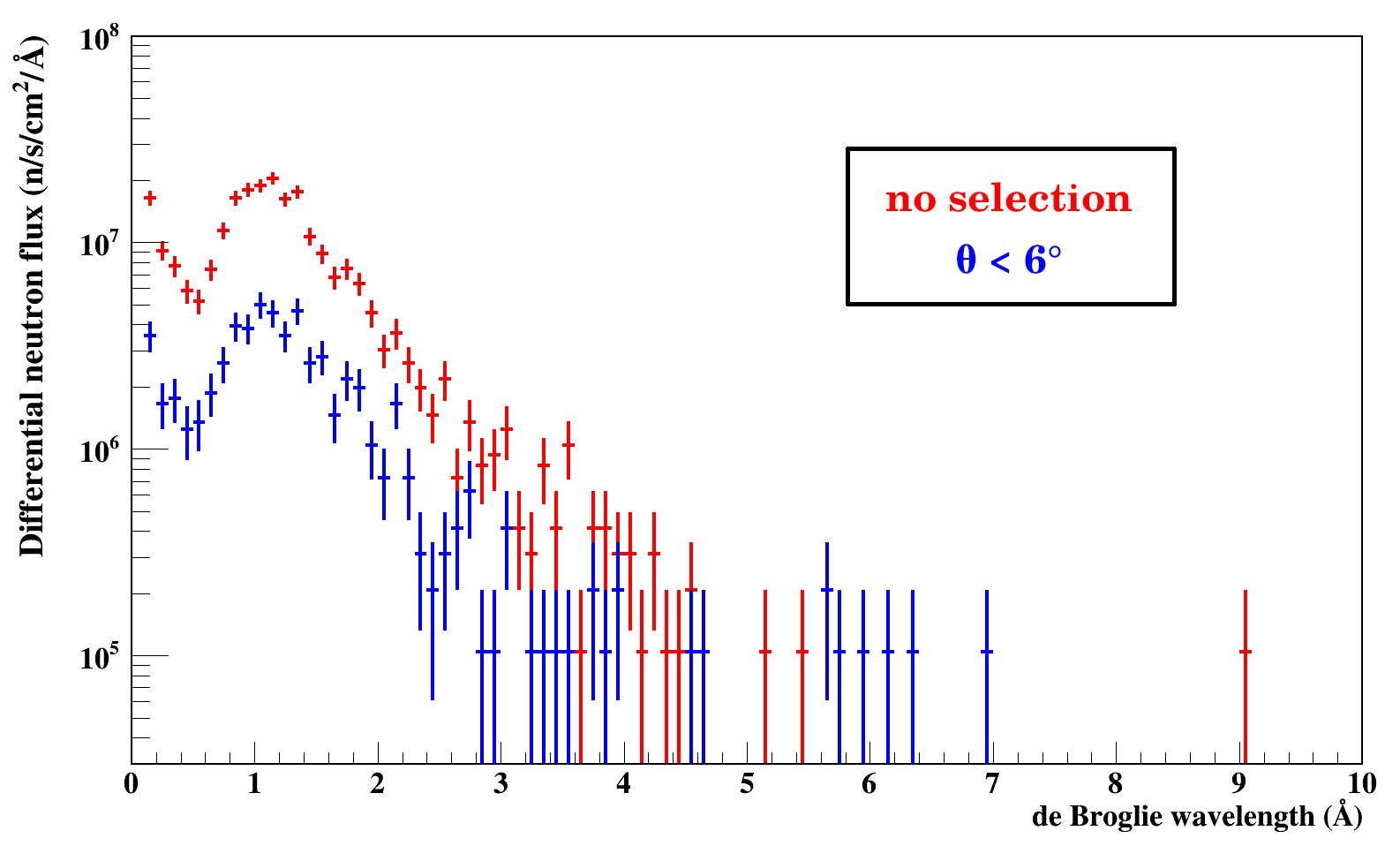}
\caption{\label{length} Neutron flux (in wavelength units) at the exit of a moderated beam extraction line for the reference moderator setup presented in \autoref{extra}.
 Red: no angle selection.
 Blue: emission angle below $6^\circ$.}
\end{figure}

\begin{figure}
\centering
\includegraphics[width=.9\columnwidth]{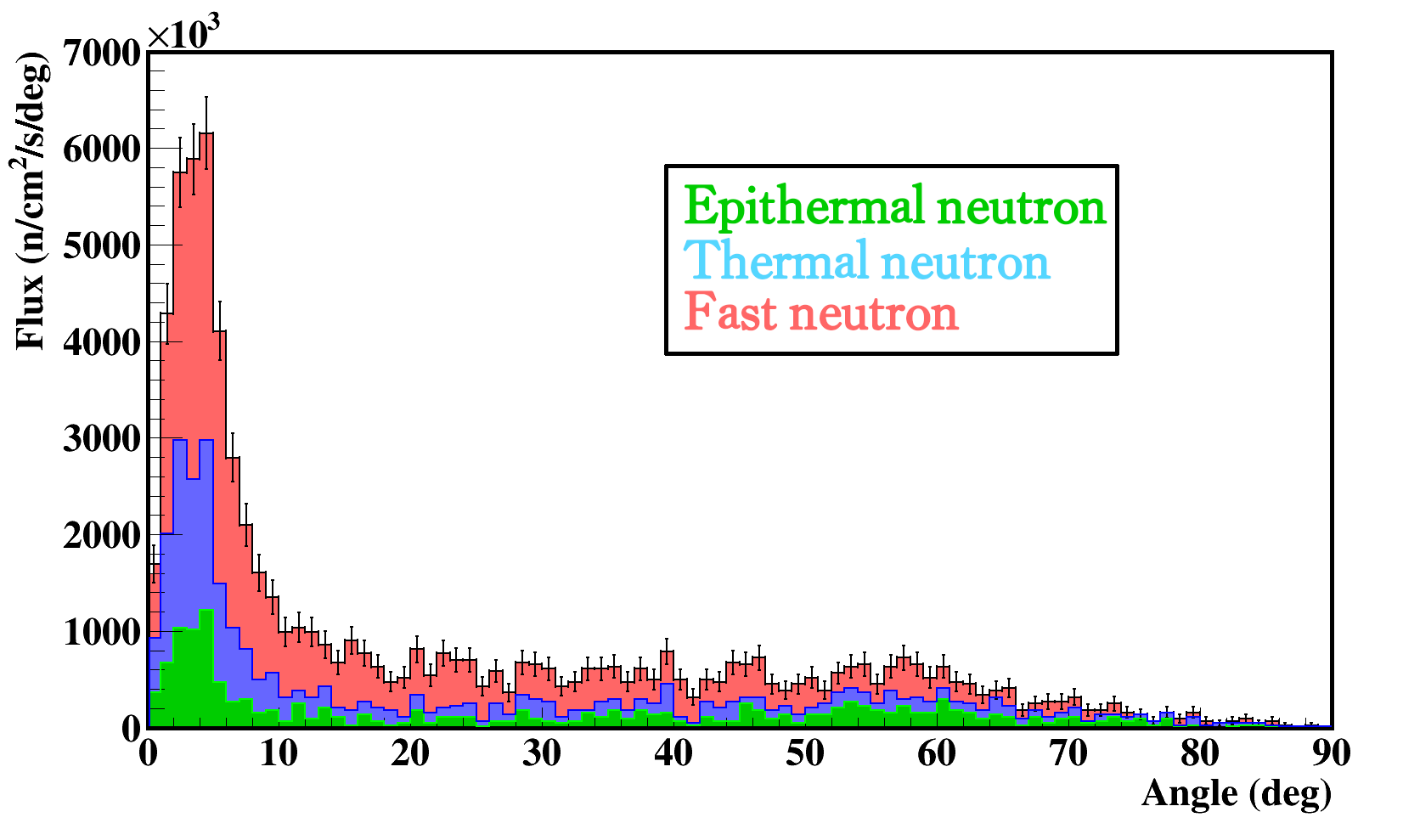}
\caption{\label{dispertion} Beam divergence at the exit of a beam extraction line for the reference moderator setup.}
\end{figure}

 For scattering experiment, the de Broglie wavelength must range between 0.5 and 40~\AA{} \cite{Fischer, d33}.
 In \autoref{length} is plotted the neutron flux extracted from the thermal moderator as a function of the de Broglie wavelength.
 With the reference setup, the neutron flux is competitive with other facilities in the range 1 to 3~\AA{} with a flux between 2~10$^7$ and 10$^6$~n/s/cm$^2$/\AA{}.
 However, the divergence of that beam is quite important as shown in \autoref{dispertion}.
 A selection in angle below 6$^\circ$ reduces the flux to 4~10$^6$~n/s/cm$^2$/\AA{} at 1~\AA{}.
 This flux is still competitive with single arm instruments for elastic scattering measurements at ILL or at small-scale facilities but it is one order of magnitude less for triple axis spectrometers \cite{d19}.

With longer extraction line necessary for imaging, we estimate a thermal neutron flux of 10$^4$~n/cm$^2$/s on a sample for a $L/D=200$ where L = 6~m is the length to the object and D = 30~mm the pin-hole diameter.
 This flux is between 2 and 3 orders of magnitude below what can be produced in experimental nuclear reactors \cite{d19}.
 
 We have to stress that this flux value is given for a beam tube with an entrance diameter of 4.5 cm.
 Increasing the entrance aperture will increase the number of neutrons available in the moderator and thus the extracted number of neutrons.
 We have studied the effect of increasing the diameter of the beam tube from 4.5 cm to 9 cm.
 It results in a factor 7 of gain on the neutron flux reaching the sample.
 Consequently, we could expect to gain one order of magnitude in the performances with a better beam tube and moderator optimization. 

\section{Summary}
\label{summary}

 In this work, we have reviewed the potential neutron fluxes which would be available at IFMIF-DONES in a dedicated hall placed behind the test cell bio-shield.
 
 The basic configuration proposed in this work consists in a beam tube crossing the test cell bio-shield of conical shape with 4.5~cm of entrance diameter and 9~cm of exit diameter.
 This simple geometry is probably not the best optimised one but it offers a simple case of reference to estimate the neutron flux inside the hall. Moreover it meets some requirements for the installation.
 With this configuration, a collimated fast-neutron beam of about 2~10$^{10}$~n/cm$^2$/s can be reached in the hall.
 Increasing the entrance diameter would increase the total number of neutrons in the hall but not the flux.
 At present, a preliminary engineering design of a neutron transport line with a shutter allowing to open and close the neutron flow is being developed as part of the activities of the Eurofusion Early Neutron Source work package
 
 To transform these energetic neutrons into thermal neutrons we have studied the possibility of adding a moderator with extraction lines, located in the hall, at the exit of the beam tube.
 With the studied configuration, the conversion efficiency is less than 0.1\% with fluxes of the order of ~10$^{6}$-10$^{7}$~n/cm$^2$/s on the instrument, with a large fast to thermal neutrons ratio that could be problematic for some experiments but could also be reduced with optimized set-up.
 In consequence, the neutron fluxes on the instruments are less than for existing state of the art instruments but not far from small-scale neutron facilities.
 We have to stress that the simulations done in this work are not optimised.
 Thus, the conclusions drawn from this work have to be taken carefully and should be used only as guidelines for further dedicated studies. 
 
 In conclusion, it appears that IFMIF-DONES has the potentialities to be a competitive multipurpose medium-flux neutron facility for most of the neutron techniques used for sample characterisation.
 Mainstream experiments, which do not require extreme neutron fluxes or high resolution, will be feasible.
 This could be helpful for preparing experiments to be performed at larger scale facilities and of interest for national uses.
 Unfortunately, in absence of deuteron beam structure, the neutron ToF technique cannot be used and consequently most of interesting nuclear physics experiments cannot be drawn.

\section*{Acknowledgments}

 The authors want to thank Grzegorz Tracz for providing the MCNP calculations of the test cell.
 This work has been supported by the European commission as part of the project IFMIF-DONES Preparatory Phase (Ref. 870186). The views and opinions expressed herein do not necessarily reflect those of the European Commission. The European Commission is not responsible for any use that may be made of it.

\end{document}